\documentclass[english,aps,twocolumn,nobalancelastpage,tightenlines,floatfix]{revtex4}

\pdfoutput=1

\usepackage[latin9]{inputenc}
\usepackage{amsmath}
\usepackage{graphicx}
\usepackage{amssymb}

\begin{document}

\title{Optical Interferometers with Reduced Sensitivity to Thermal Noise}

\author{H.~J. Kimble$^{\dagger}$, Benjamin L. Lev$^{\ddag}$, and Jun Ye}

\affiliation{JILA, National Institute of Standards and Technology and University
of Colorado, Boulder, CO \ 80309--0440}

\begin{abstract}
A fundamental limit to the sensitivity of optical interferometry is
thermal noise that drives fluctuations in the positions of the surfaces
of the interferometer's mirrors, and thereby in the phase of the intracavity
field. Schemes for reducing this thermally driven phase noise are
presented in which phase shifts from concomitant strains at the surface
and in the bulk of the substrate compensate the phase shift due to
the displacement of the surface. Although the position of the physical
surface fluctuates, the optical phase upon reflection can have reduced
sensitivity to this motion. 
\end{abstract}

\date{\today}

\maketitle
Thermal noise presents a fundamental limit to measurement sensitivity
in diverse areas of science and technology~\cite{braginsky86}.
One important setting is that of optical interferometry in which otherwise
stable structures experience small, thermally driven fluctuations
in their dimensions in applications ranging from frequency metrology~\cite{numata03,numata04,ludlow07},
to gravitational wave detection~\cite{barish99,rowan05}, to the
realization of quantum behavior for macroscopic objects~\cite{q-mechanical}.

The dominant limitation to length stability for some interferometers
originates from thermally driven displacement noise for the reflective
surfaces of the mirror substrates, and not from the external, support
structure~\cite{rowan05,numata04,ludlow07}. The fluctuations of
the mirror surfaces are of fundamental origin and arise from dissipation
in the elasticity of the substrate as demanded by the Fluctuation-Dissipation
Theorem (FDT)~\cite{callen52,greene52,LL}.

Beyond displacements driven by thermal noise in the substrate itself~\cite{saulson90,gillespie95,nakagawa97,levin98,bondu98,liu00},
diverse other sources of mechanical noise have been identified in
recent years, including noise from frictional losses in the materials
that form the mirror coating~\cite{levin98,harry02}, thermoelastic-damping
in the substrate and coating~\cite{braginsky99,liu00,braginsky03},
and thermo-refractive noise \cite{braginsky00}. Measurements of
phase noise in rigid~\cite{numata03,numata04,ludlow07}\ and suspended
interferometers~\cite{black04a,black04b,harry07} have confirmed
many characteristics of individual noise sources.

Various avenues have been followed for reducing thermal noise in optical
interferometers, the most significant being lowering mechanical losses
for the substrate~\cite{braginsky86,penn06} and, more recently,
the coating~\cite{fejer04,harry07}. New designs for advanced interferometers
include the use of a corner reflector~\cite{braginsky04} or a
short Fabry-Perot (FP) cavity~\cite{khalili05} to replace the usual
single surface of a mirror.

In this Letter, we propose a new strategy for reducing thermally driven
phase noise in optical interferometers. Fundamental to our proposal
is the observation that stochastic displacements $\delta u_{z}$ perpendicular
to the mirror's surface are necessarily accompanied by correlated
strains in the underlying materials of the mirror coating and substrate.
In a conventional setting, the phase shift $\delta\beta$ due to these
strains in the coating and substrate are small compared to the phase
shift $\delta\theta$ from the surface motion $\delta u_{z}$. However,
by suitable design, it should be possible to achieve a total phase
shift $\delta\Phi=\delta\beta+\delta\theta\simeq0$ for the reflected
field. That is, although the physical surface of the mirror is subject
to random displacements $\delta u_{z}$, $\delta\Phi$ can be insensitive
to these displacements with $\delta\theta$ dynamically compensated
by $\delta\beta$ from the coating and substrate. Although our current
analysis is for {}``Brownian noise'' of the substrate, our methodology
should be applicable to thermoelastic and coating noise as well.

As a first example, consider thermal noise for an eigenmode of a cylindrical
mirror with mechanical resonance frequency $\omega_{0}$. For frequencies
$\omega\simeq\omega_{0}$ and quality factor $Q_{0}\gg1$, microscopic
thermal noise excites the entire {}``shape'' of the relevant eigenmode
$\boldsymbol{\xi}_{0}(\boldsymbol{r})$, with a single, overall amplitude
set by equipartition of energy~\cite{saulson90,gillespie95}. The
amplitude $\delta u_{z}$ and accompanying strain $\epsilon_{zz}$
for thermally driven motion perpendicular to the mirror surface at
$z=0$ are then determined directly from ${\boldsymbol{\xi}_{0}}(\boldsymbol{r})$.

\begin{figure}[tb]
 \includegraphics[width=8.6cm]{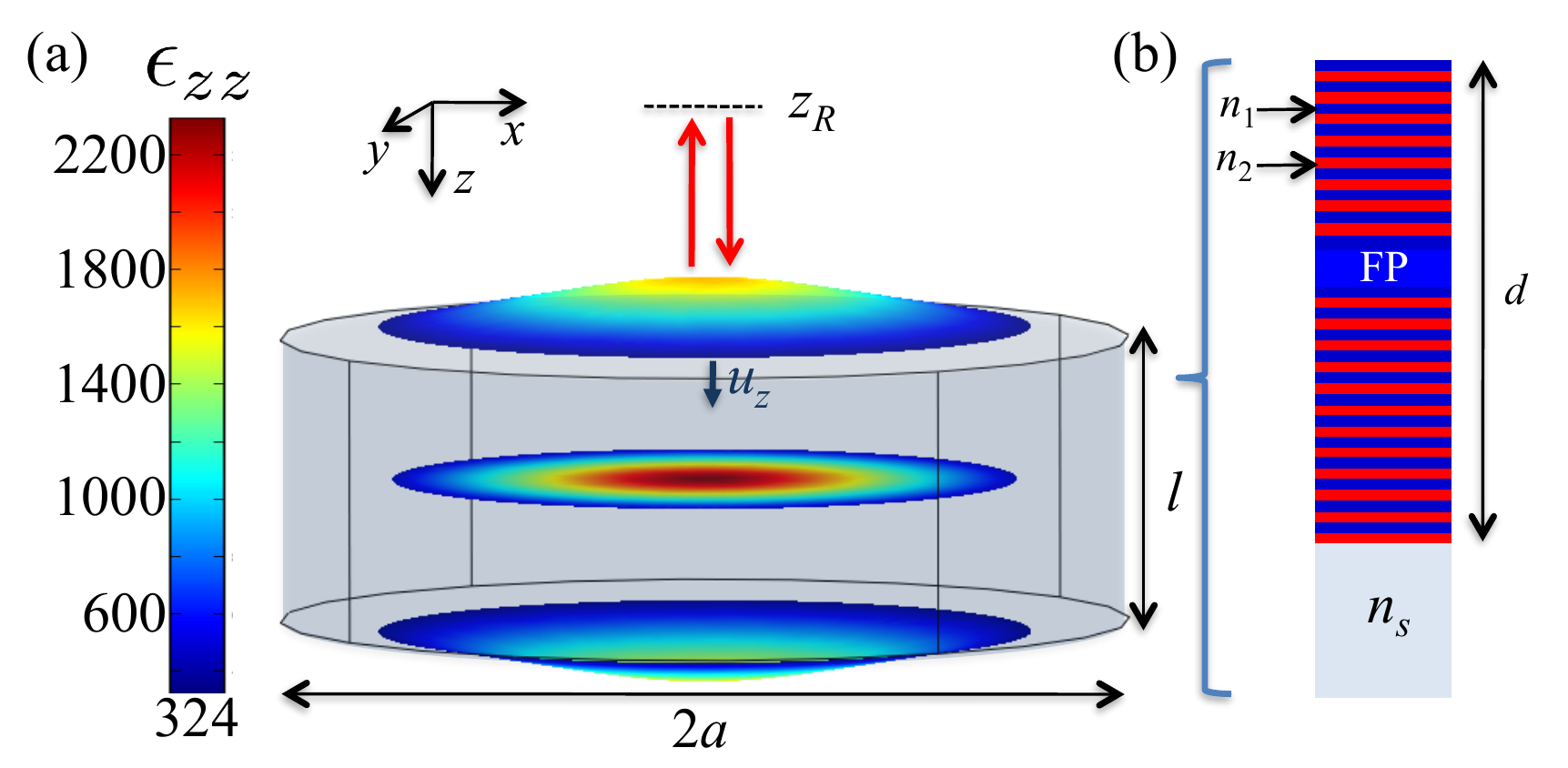} 
\caption{(color online) (a) The undistorted shape of a sapphire substrate of
radius $a=1.5$ mm and thickness $l=1$ mm is depicted by the shaded
region and wire frame. Axial displacements $u_{z}(x,y,z)$ and strains
$\epsilon_{zz}(x,y,z)$ are shown across planes at $z=0,l/2,l$ (top
to bottom) for the eigenmode $\boldsymbol{\xi}_{0}(\boldsymbol{r})$
with eigenfrequency $\omega_{0}/2\pi=2.22$ MHz. Plotted are contours
for $u_{z}(x,y,z)$ on which $\epsilon_{zz}(x,y,z)$ is color coded
at a time of maximum axial extension, where $u_{z}(x,y,z)=0$ absent
excitation. The phase for propagation from $z=z_{R}$ to the substrate
and back is modified by surface motion $u_{z}$ at $z=0$. (b) Coating
stack for a high reflectivity mirror with embedded FP cavity for high
strain sensitivity.}
\label{sapphire} 
\end{figure}

Figure~\ref{sapphire}(a) illustrates a particular axisymmetric eigenmode
$\boldsymbol{\xi}_{0}(\boldsymbol{r})$ for a sapphire substrate of
mass $M$ as determined from a numerical finite-element analysis.
For this mode, the end faces at $z=0,l$ oscillate in opposition about
the plane at $z=l/2$ with frequency $\omega_{0}$. In thermal equilibrium
at temperature $T$, $\delta u_{z}$ at the central points $x,y=0$
on the end faces has amplitude $\langle\delta u_{z}^{2}\rangle_{\omega_{0}}\simeq k_{B}T/M_{0}\omega_{0}^{2}$,
where $M_{0}$ is the effective mass~\cite{gillespie95}, which
for the mode shown is $M_{0}=2.7M$. Significant to our current investigation
is the axial strain per unit displacement, $\zeta\equiv\epsilon_{zz}/u_{z}$,
where $\zeta\simeq-1600/$m at $x,y,z=0$ for the parameters in Fig.
\ref{sapphire}(a) (i.e., expansion at $z=0$ with $u_{z}<0$ generates
strain $\epsilon_{zz}>0$).

As depicted in Fig.~\ref{sapphire}, we next address the question
of the phase shift for light reflected from the fluctuating surface
of the substrate at $z=0$. Following Levin~\cite{levin98}, we
introduce the displacement $q(z)$ weighted by the normalized light
intensity $\psi(r)$ over a plane at depth $z\geq0$, namely
$q(z)=\int dxdy\delta u_{z}(r,z)\psi(r)\text{,}$ where $\psi(r)=(2/\pi w_{0}^{2})\exp(-2r^{2}/w_{0}^{2})$ with $r=\sqrt{x^{2}+y^{2}}$.
The incident field with vacuum wavevector $k$ experiences a phase
change $\delta\theta=-2kq(z=0)\equiv-2kq_{0}$ for the reflected field
due to the {}``piston'' motion of the surface.

In addition to $\delta\theta$, strain that accompanies surface motion
modifies the optical coating and thereby leads to a phase shift $\delta\beta$,
with $\delta\beta$ expressed relative to the front surface of the
coating. The overall phase shift for the reflected field is then $\delta\Phi=\delta\theta+\delta\beta$,
with $|\delta\beta|\ll|\delta\theta|$ for typical optical coatings.
We now present new designs for the coating to achieve $\delta\Phi=\delta\theta+\delta\beta\simeq \delta\theta/Q_{0}$.

The coating structure shown in Fig.~\ref{sapphire}(b) has an embedded
resonant layer that functions as a FP cavity and gives rise to a rapid
phase variation near the cavity resonance~\cite{mcleod}. Explicitly,
the coating structure is specified as $n_{0}(\eta_{1}\eta_{2})^{l}\eta_{1}(j\times\eta_{FP})(\eta_{1}\eta_{2})^{p-l}n_{s}$.
Starting from the vacuum side $n_{0}=1$, there are $l$ double-layers
$(\eta_{1}\eta_{2})^{l}$ with $\eta_{1,2}=\pi/2$ at the reference
wavevector $k_{0}$, followed by an $n_{1}$ layer with $\eta_{1}=\pi/2$
at $k_{0}$, then the FP $n_{1}$\ layer with single-pass phase shift
$j\times\eta_{FP}$ at $k_{0}$, followed by the terminating $p-l$
double layers, and lastly the substrate with index $n_{s}$. We restrict
attention to the case of a {}``thin'' coating for which the total
coating thickness $d\ll a,w_{0}$ and further assume that the axial
strain in the coating $\epsilon_{zz}^{c}$ is the same as that in
the substrate $\epsilon_{zz}$. These assumptions simplify the discussion
of the essential aspects of our scheme for noise compensation; see
Refs.~\cite{harry07,fejer04} for detailed treatments of multilayer
coatings. A discussion of thermal noise from the coating itself is
deferred to our concluding remarks.

\begin{figure}[t]
\includegraphics[width=8.6cm]{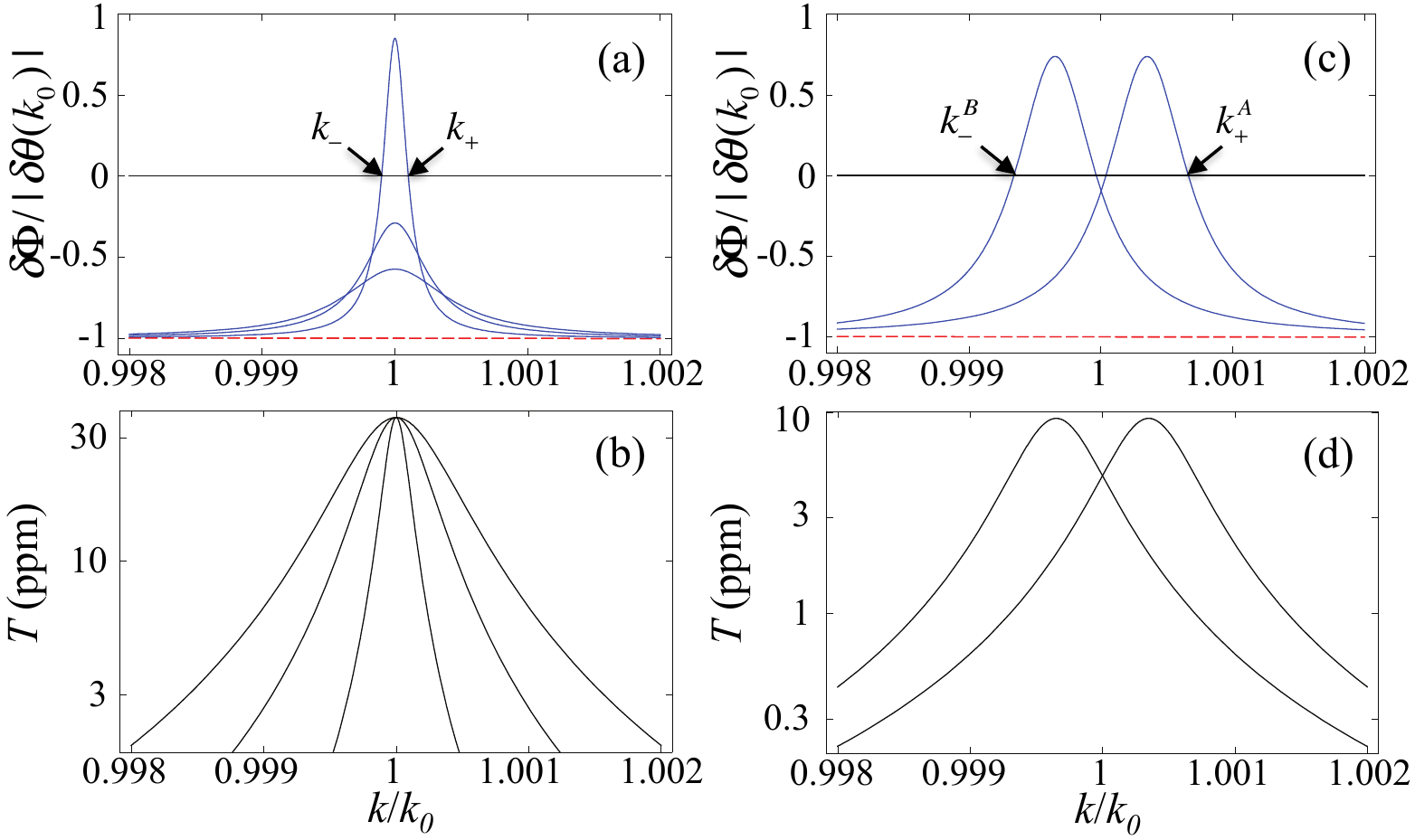} 
\caption{Phase shift $\delta\Phi(k)$ and transmission coefficient $T(k)$
for two different coating designs illustrated in Fig.~\ref{sapphire}(b)
and specified in the text. (a, b) The case $\zeta=-1600$/m for increasing
phase $j\eta_{FP}$ of the internal resonant structure. (a) Total
phase $\delta\Phi(k)$ with {}``magic'' wavevectors $k_{\pm}$ and
(b) associated transmission coefficient $T(k)$ for $j=1,4,16$ from
widest to narrowest curves. (c, d) $\delta\Phi(k)$ and $T(k)$ for
the case $\zeta=-5000/$m for two coatings with slightly different
internal resonances $\eta_{FP}^{(A,B)}$ with distinct values $k_{-}^{B},k_{+}^{A}$
for nulling noise from mirrors $B,A$. In all cases, $\delta\Phi(k)/|\delta\theta(k_{0})|$
is plotted for $q_{0}>0$, with $\delta\theta(k)$ given by the dashed
lines in (a, c).}
\label{2FPcoatings} 
\end{figure}

Results for two particular coatings are given in Fig.~\ref{2FPcoatings}
for $w_{0}\ll a$. For definiteness, we assume a coating stack made
from layers of \emph{SiO$_{2}$} with index $n_{1}=1.45$ and \emph{Ta$_{2}$O$_{5}$}
with $n_{2}=2.03$. Parts (a, b) are for displacement-induced strain
$\zeta=-1600/$m, as appropriate to $\boldsymbol{\xi}_{0}(\boldsymbol{r})$
in Fig.~\ref{sapphire}(a). The coating structure is
$\left[p=33,l=8\right]$ with three curves drawn for mode orders $j=1,4,16$
with $\eta_{FP}=\pi$. In (a), increased length for the resonant structure
leads to greater strain sensitivity, with $|\delta\beta|$ becoming
larger relative to $|\delta\theta|$. For $j=16$ there arise {}``magic''
wavevectors $k_{\pm}$ for which $\delta\Phi(k_{\pm})=0$, with then
the piston phase shift $\delta\theta$ from fluctuations of the surface
dynamically compensated by the strain-induced phase shift $\delta\beta$
from the coating. At these {}``magic wavelengths,'' the phase of
the reflected optical field becomes insensitive to the thermal motion
of the surface of the substrate.

Figures~\ref{2FPcoatings}(c,d) investigate the possibility that
thermal noise of individual mirrors of an interferometer could be
measured \textit{in situ}. The coating for each of two mirrors $A,B$
is specified by $\left[p=33,l=7,j=8\right]$. The thicknesses of the FP layers in
the $A,B$ coatings have been adjusted to give $\eta_{FP}^{(A,B)}=(0.9995,1.0005)\times\pi$
at $k_{0}$ for the two traces shown. For operation at the lower (upper)
value $k_{-}^{B}(k_{+}^{A})$ noise from mirror $B(A)$ would be nulled
while that from mirror $A(B)$ would be $\sim80\%$ of the full piston
phase, so that the noise arising from mirrors $A,B$ could be individually
measured. For operation at the central value $k/k_{0}=1$, thermal
noise from both mirrors $A,B$ would be suppressed. Panel (d) gives
the transmission coefficient $T(k)=1-R(k)$ for the two coatings,
where $T(k_{-}^{B},k_{+}^{A})$ is consistent with high-finesse measurements.
Not shown in Fig.~\ref{2FPcoatings} are FP resonances away from
$k\approx k_{0}$, which can also be employed to suppress thermal
noise.

Because the surface strain $|\zeta|$ decreases with increasing size
of the substrate, compensation of the piston phase $\delta\theta$
with increasing $a,w_{0}$ requires greater departures from standard
coating designs, material specifications, and fabrication procedures.
Moreover, not all eigenmodes of oscillation have the proper parity
(i.e., $\zeta<0$) for compensation by way of the coating designs
in Fig.~\ref{2FPcoatings}.

A second, more challenging example is thermal noise at frequencies
$\omega$ well below the lowest resonance of the mirror substrate.
In this quasi-static regime, we must understand the correlation between
thermal displacements at depth $z$ and fluctuations of the surface
arising from many modes~\cite{gillespie95}. The foundation for
our analysis is the FDT \cite{callen52,greene52} as applied by Levin
to this setting~\cite{levin98}.

\begin{figure}[tb]
\includegraphics[width=8.6cm]{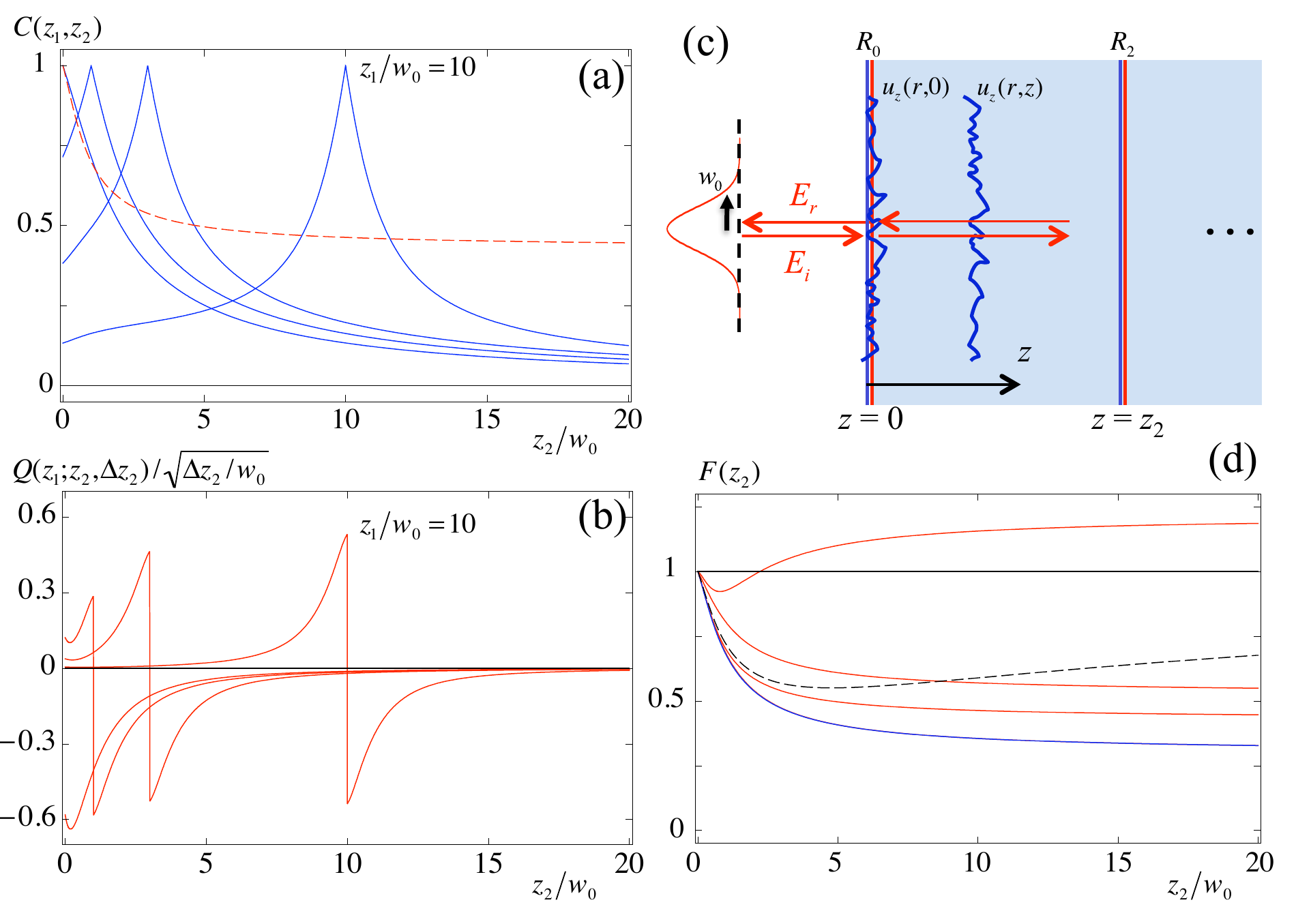} 
\caption{(a) Correlation of thermally driven axial displacements $C(z_{1},z_{2})$
and (b) displacement-strain correlation $Q(z_{1};z_{2},\Delta z_{2})/\sqrt{\Delta z_{2}/w_{0}}$
versus $z_{2}/w_{0}$ for $z_{1}/w_{0}=0,1,3,10$, with $\Delta z_{2}/w_{0}=10^{-3}$
in (b). The dashed trace in (a) is $N(z_{2},z_{2})$. (c) Illustration
of the mirror geometry discussed in the text. (d) Spectral density
of phase fluctuations $\digamma(z_{2})$ for the reflected field $\boldsymbol{E_{r}}$
for the configuration in (c) for an $SiO_{2}$ substrate. From top
to bottom, the curves are for $\alpha=1.5,0.3,1.0,0.7$ with the blue
trace for $\alpha=\alpha_{min}$ overlaying the curve for $\alpha=0.7$.
The dashed curve is from a simple model that incorporates incoherent contributions
from transverse strains to $\delta\beta$. All curves are for $\sigma=0.2$.}
\label{CQF} 
\end{figure}

We consider a substrate in the form of an infinite half space with
boundary at $z=0$ and extending to $z\geq0$. Equation~(9) in Mindlin~\cite{mindlin36}
provides the Green's function required to deduce the admittance for
application of the FDT from Eq.~(6.8) of Ref.~\cite{greene52}.
Following the techniques in~\cite{liu00}, we find the spectral
correlation for beam-averaged axial displacements $\widetilde{q}(z)$
at depths $z_{1},z_{2}$ to be \begin{equation}
\langle\widetilde{q}(z_{2})\widetilde{q}(z_{1})\rangle_{\omega}=\langle\widetilde{q}_{0}\widetilde{q}_{0}\rangle_{\omega}N(z_{2},z_{1})\text{,}\label{qq}\end{equation}
 where $\widetilde{q}(z_{i})\equiv\widetilde{q}(z_{i},\omega)$ is the
Fourier transform of $q(z_{i},t)$. $\langle\widetilde{q}_{0}\widetilde{q}_{0}\rangle_{\omega}=2k_{B}T(1-\sigma^{2})\phi_{s}/\pi^{3/2}w_{0}E\omega$
is the standard result for thermally driven surface displacements,
with $E,\sigma$ the Young's modulus and Poisson ratio, respectively,
and $\phi_{s}$\ the (possibly frequency dependent) loss angle for
the substrate. The function $N(z_{2},z_{1})$ determines the mechanical admittance
and is given by
\begin{equation}
\lefteqn{N(z_{2},z_{1})=\frac{w_{0}}{8\sqrt{\pi}(1-\sigma)^{2}}\int_{0}^{\infty}dke^{-k^{2}w_{0}^{2}/4}f(z_{2},z_{1};k),} \ \ \ \ \ \ \ \ \ \ \ \ \ \ \ \ \ \ \ \ \ \ \ \ \ \ \ \ \ \ \ \ \ \ \ \ \ \ \ \ \ \ \ \ \ \ \ \ \ \ \ \ \ \ \ \ \ \ \ \ \ \ \ \ \ \ \ \ \ \ \ \ \ \ \ \ \ \ \ \ \ \ \ \ \nonumber
\end{equation}
where
\begin{eqnarray}
\lefteqn{f(z_{2},z_{1};k)=e^{-k|z_-|}\left[3-4\sigma+k|z_-|\right]+} \ \ \ \ \ \ \ \ \ \ \ \ \ \ \ \ \ \ \ \ \ \ \ \ \ \ \ \ \ \ \ \ \ \ \ \ \ \ \ \ \ \ \ \ \ \ \ \ \ \ \ \ \ \ \ \ \ \ \ \ \ \ \ \ \ \ \ \ \ \ \nonumber\\
\lefteqn{e^{-kz_+}\left[5-12\sigma+8\sigma^{2}+k(3-4\sigma)z_++2k^{2}z_{1}z_{2}\right],} \ \ \ \ \ \ \ \ \ \ \ \ \ \ \ \ \ \ \ \ \ \ \ \ \ \ \ \ \ \ \ \ \ \ \ \ \ \ \ \ \ \ \ \ \ \ \ \ \ \ \ \ \ \ \ \ \ \ \ \ \ \ \ \ \ 
\end{eqnarray}
with $z_{\pm} = z_1 \pm z_2$ and $N(z,z)$ plotted in Fig.~\ref{CQF}(a). Note that $N(0,0)=1$,
$N(z_{1},z_{2})=N(z_{2},z_{1})$ \cite{LL}, and that the admittance
derived from $N(z,0)$ agrees with that from previous work~\cite{levin98,liu00,harry02}.

Equation~(\ref{qq}) enables us to determine spectral correlations
between thermally driven axial displacements at depths $z_{1},z_{2}$
characterized by~\cite{brevity} \begin{equation}
C(z_{1},z_{2})=\frac{\langle\widetilde{q}(z_{1})\widetilde{q}(z_{2})\rangle}{\langle\widetilde{q}(z_{1})\widetilde{q}(z_{1})\rangle^{1/2}\langle\widetilde{q}(z_{2})\widetilde{q}(z_{2})\rangle^{1/2}}\text{,}\label{Cz1z2}\end{equation}
 where $C(z,z)=1$ and $|C(z_{1},z_{2})|\leq1$. From Fig.~\ref{CQF}(a),
we see that thermal fluctuations of $\widetilde{q}(z)$ correlate
over length scales set by $w_{0}$, albeit with decreasing amplitude
$\langle\widetilde{q}(z)\widetilde{q}(z)\rangle\varpropto N(z,z)$
away from the surface.

Figure~\ref{CQF}(b) investigates correlation between displacement
$\widetilde{q}(z_{1})$ at $z_{1}$ and axial strain at depth
$z_{2}$ by way of the function $Q(z_{1};z_{2},\Delta z_{2})=\varepsilon_{z}^{coh}(z_{1},z_{2})/\varepsilon_{z}^{tot}(z_{2},\Delta z_{2})$,
with $|Q|\leq1$. $Q(z_{1};z_{2},\Delta z_{2})$ expresses the ratio
of coherent to total strain at $z_{2}$ within a small slice $\Delta z_{2}$,
where $\varepsilon_{z}^{coh}(z_{1},z_{2})\equiv\langle\widetilde{q}(z_{1})[\widetilde{q}(z_{2}+\Delta z_{2})-\widetilde{q}(z_{2})]/\Delta z_{2}\rangle/\langle\widetilde{q}(z_{1})^{2}\rangle^{1/2}$
is the strain at $z_{2}$ correlated with the displacement $\widetilde{q}(z_{1})$
at $z_{1}$, while $\varepsilon_{z}^{tot}(z_{2},\Delta z_{2})\equiv\langle[\widetilde{q}(z_{2}+\Delta z_{2})-\widetilde{q}(z_{2})]^{2}\rangle^{1/2}/\Delta z_{2}$
is the total strain at $z_{2}$.

In Fig.~\ref{CQF}(b), the spatial scale for correlation of displacement
and strain is again set by $w_{0}$, but now with magnitude reduced
by $\sqrt{\Delta z_{2}/w_{0}}$. This scaling of $Q$ can be understood
from the fact that the rms strain diverges as $1/\sqrt{V}$ for thermal
fluctuations in a volume $V$~\cite{parrinello82}, which motivates
our use of finite differences to characterize strain. Near the surface
at $z=0$ with $\Delta z/w_{0}\ll1$, we find from Eq.~(\ref{qq})
that $\varepsilon_{z}^{tot}(z\simeq0,\Delta z)\sim[\langle\widetilde{q}_{0}\widetilde{q}_{0}\rangle/(\Delta zw_{0})]^{1/2}$,
where the relevant volume is $V\sim\pi w_{0}^{2}\Delta z$. On the
other hand, the coherent strain $\varepsilon_{z}^{coh}\equiv\varepsilon_{z}^{coh}(0,0)\sim-\langle\widetilde{q}_{0}\widetilde{q}_{0}\rangle^{1/2}/w_{0}$,
so that $|\varepsilon_{z}^{coh}/\varepsilon_{z}^{tot}|$ scales as
$\sqrt{\Delta z/w_{0}}\ll1$ near $z=0$.

Simply stated, thermally driven surface motion $\widetilde{q}_{0}$
arises from strains within the substrate over distances $z\gtrsim w_{0}$.
Hence, our previous strategy with a thin optical coating of thickness
$\Delta z=d\ll w_{0}$ employed as a surface-strain sensor for a single
eigenmode will compenstate only a small fraction of the phase shift
$\delta\theta=-2k\widetilde{q}_{0}$ from the surface motion in the
quasi-static regime. Instead, we must find a geometry for which $\delta\beta$
has contributions over $z\gtrsim w_{0}$ sufficient to compensate
$\delta\theta$.

An initial mirror geometry that attempts to achieve such compensation
is illustrated in Fig.~\ref{CQF}(c). A thin dielectric coating with
reflectivity $R_{0}$ is deposited on the surface of the substrate
at $z=0$, and a second reflecting surface is embedded at $z=z_{2}\ll nk{w_{0}}^{2}/2$
with reflectivity $R_{2}\rightarrow1$. In the static case, an incident
field $\boldsymbol{E_{i}}$ linearly polarized along $x$ is reflected
from this two-mirror geometry to give a field $\boldsymbol{E_{r}}$
with $\boldsymbol{E_{r}}/\boldsymbol{E_{i}}=Ae^{i\Gamma(\varphi)}$,
where $\varphi$ is the roundtrip, internal phase, and $A=1$ for
$R_{2}=1$. Thermally driven fluctuations lead to a phase shift $\delta\Phi$
for $\boldsymbol{E_{r}}$, $\delta\Phi=-2k\left[\widetilde{q}_{0}+\alpha(k)\left(\widetilde{q}(z_{2})-\widetilde{q}_{0}\right)\right],$
where the piston phase is $\delta\theta=-2k\widetilde{q}_{0}$ and
the interferometer phase is $\delta\beta=-2k\alpha(k)\left(\widetilde{q}(z_{2})-\widetilde{q}_{0}\right)$.
$\alpha(k)$ expresses the sensitivity of the composite cavity $R_{0},R_{2}$
to strain driven changes in both length and index around $\varphi_{0}$,
with \begin{equation}
\alpha(k)=n_{s}\left(1-\frac{n_{s}^{2}p_{12}}{2}\right)\left|\frac{d\Gamma(\varphi_{0})}{d\varphi}\right|,\label{alpha}\end{equation}
 where $p_{12}$ is an element of the strain-optic tensor $p_{ij}$
for an assumed isotropic material. In Eq.~\ref{alpha}, we neglect
contributions from fluctuating strains transverse to the propagation
direction $z$, which are considered below.

The expressions for $\delta\Phi$ and $\alpha(k)$ lead to the spectral
density $\left\langle (\delta\Phi)^{2}\right\rangle$ of phase fluctuations
for the reflected field $\boldsymbol{E_{r}}$, namely $\left\langle (\delta\Phi)^{2}\right\rangle =\left\langle (\delta\theta)^{2}\right\rangle [\left(1-\alpha\right)^{2}+2\alpha\left(1-\alpha\right)N(0,z_{2})+\alpha^{2}N(z_{2},z_{2})].$
Figure~\ref{CQF}(d) plots $\digamma(z_{2})\equiv\left\langle (\delta\Phi)^{2}\right\rangle /\left\langle (\delta\theta)^{2}\right\rangle $
for several values of $\alpha$. For fixed $R_{0}$ (with $R_{2}=1$),
$\alpha(k)$ varies periodically from minimum to maximum over the
range $\Delta k=\pi/n_{s}z_{2}$, with then $\alpha$ determined by
the selection of $k$. $\digamma<1$ represents phase noise reduced
below that from thermal fluctuations in the piston phase $\left\langle (\delta\theta)^{2}\right\rangle $.
Since $\delta\Phi=2k\left[(1-\alpha)\widetilde{q}_{0}+\alpha\widetilde{q}(z_{2})\right],$
$\alpha=1$ corrresponds to direct compensation of the piston
phase $\delta\theta\propto\widetilde{q}_{0}$ for the reflected field,
albeit at the price of noise $\widetilde{q}(z_{2})$ from fluctuations
in the position of the reflecting surface at $z_{2}$. More generally,
the minimum $\digamma_{min}$ for a given value of $z_{2}$ represents
a compromise in noise from $\widetilde{q}_{0}$ and $q(z_{2})$ determined
by the value $\alpha_{min}$. $\digamma_{min}\simeq0.36$ for $z_{2}\gg w_{0}$
and $\sigma=0.2$ in Fig.~\ref{CQF}(d).

An important caveat related to Fig.~\ref{CQF}(d) is that the full
curves omit fluctuations in optical path arising from transverse strains
$\epsilon_{xx},\epsilon_{yy}$, which contribute by way
of $p_{ij}$ to $\delta\beta$ and give rise to a scaling $\digamma\sim z_{2}/w_{0}$
for $z_{2}/w_{0}\gg1$. A full treatment of these effects is beyond
the scope of our current analysis. Instead, the dashed curve in Fig.
\ref{CQF}(d) is from a simple model based upon the FD theorem applied to $\epsilon_{xx},\epsilon_{yy}$  with loss angle $\phi_{s}$ and
provides a rough estimate of their incoherent contribution
to $\delta\beta$.

The conceptual design in Fig.~\ref{CQF}(c) is likely far from optimal.
Because the strain field associated with $\psi(r)$ at $z=0$ spreads
transversely for $z>0$, it is not well matched to our assumed optical
profile with fixed $w(z)=w_{0}$. Geometries with partially reflecting
surfaces distributed along $z$ might further reduce $\digamma$.
For finite thickness of the substrate, a treatment as in~\cite{liu00}
is required, with now the possibility of reflection from the rear
surface \cite{khalili05}. More generally, coherent measurements
over a range of $k$ values could enhance sensitivity since $\delta\beta(k)$
can be tailored to be distinct from $\delta\theta(k)$.

Although our treatment has been exclusively for {}``Brownian'' noise
arising in the substrate, we suggest that our methods should be relevant
to the suppression of thermal fluctuations from other sources within
the substrate, such as thermoelastic-damping~\cite{braginsky99,liu00,braginsky03}.
Moreover, variations in the coating design from Fig.~\ref{2FPcoatings}
could lead to schemes for suppression of thermal noise within the
coating~\cite{harry07}. In contrast to the substrate for which
strains at $z\ll w_{0}$ have small correlation with $\widetilde{q}_{0}$
in the quasistatic regime, thermal noise from the coating leads to
surface strains that are highly correlated with $\widetilde{q}_{0}$.

Certainly, important questions remain related to our proposals for
noise compensation, including significant fabrication challenges,
the impact of optical absorption within the coating and substrate,
and the need for more complete theoretical analyses. We make no claim
of a {}``magic bullet'' for the elimination of thermally driven
phase fluctuations in optical interferometry. Rather, our goal is
to provide a perspective on thermal fluctuations which moves beyond
a surface-centric view to consider the statistical character of the
underlying stochastic displacements and strains that conspire to displace
the surface and thereby to suggest new strategies for enhanced sensitivity
and stability of optical interferometers.

We gratefully acknowledge the guiding hand and critical insights of
K. S. Thorne, as well as valuable discussions with V. B. Braginsky,
Y. Chen, J. L. Hall, R. Lalezari, and D. R. Nelson. The work of HJK
was made possible as a Visiting Fellow at JILA. This research is supported
by the NSF and NRC.

{\scriptsize Permanent addresses: $^{\dagger}$HJK - Norman Bridge
Laboratory of Physics 12-33, California Institute of Technology, Pasadena,
CA 91125; $^{\ddag}$BLL - Department of Physics, University of Illinois,
1110 West Green St., Urbana, IL 61801}{\scriptsize \par}


\begin{thebibliography}{10}
\bibitem{brevity} For brevity, henceforth we drop the $\omega$ subscript
for $\left\langle \right\rangle _{\omega}$.
\end{thebibliography}
\end{document}